\documentclass[12pt,preprint]{aastex}

\begin{document}

\def\pdot {\dot P}
\def\Omdot {\dot \Omega}
\def\ltsima{$\; \buildrel < \over \sim \;$}
\def\lsim{\lower.5ex\hbox{\ltsima}}
\def\gtsima{$\; \buildrel > \over \sim \;$}
\def\gsim{\lower.5ex\hbox{\gtsima}}
\def\msole{~M_{\odot}}
\def\mdot {\dot M}
\def\rxj {RX~J1856.5--3754}
\def\cha {\textit{Chandra~}}
\def\xmm  {\textit{XMM-Newton~}}
\slugcomment{Accepted for publication on The Astroph. Journal
Letters}

\title{XMM-Newton discovery of 7 s pulsations in the isolated neutron star \rxj\ }
\author{Andrea Tiengo \& Sandro Mereghetti}
\affil{INAF - Istituto di Astrofisica Spaziale e Fisica Cosmica - Milano, \\
v.Bassini 15, I-20133 Milano, Italy \\
tiengo@iasf-milano.inaf.it, sandro@iasf-milano.inaf.it}

\begin{abstract}

Thanks to the high counting statistics provided by a recent
XMM-Newton observation of \rxj\, we have discovered that this
isolated neutron star pulsates at a period of 7.055 s.  This
confirms that \rxj\ is similar in nature to the other six
thermally emitting, nearby neutron stars discovered in soft
X--rays with ROSAT. The pulsations are detected at consistent
periods in several XMM-Newton observations spanning from April
2002 to October 2006, yielding an upper limit of
$\pdot<1.9\times10^{-12}$ s s$^{-1}$ (90\% c.l.) on the period
derivative. This implies a surface magnetic field smaller than
1.2$\times10^{14}$ G, under the usual assumption of vacuum dipole
magnetic braking.  The pulse profile is nearly sinusoidal with a
pulsed fraction in the 0.15-1.2 keV range of only
$\sim$1.2\%, the smallest ever seen in an isolated X-ray pulsar.

\end{abstract}


\section{Introduction}

The soft X--ray source \rxj\ belongs to a small group of seven
nearby, isolated neutron stars  discovered with the ROSAT
satellite and often dubbed X--ray Dim Isolated Neutron Stars
(XDINS,  see \citet{haberl2006} for a recent review).  These
sources are characterized by thermal spectra with blackbody
temperature in the range 40--110 eV, X--ray luminosity of
10$^{30}$--10$^{32}$ erg s$^{-1}$, faint optical counterparts
(V$>$25), and absence of radio emission. The initial suggestion
that XDINS could be powered by accretion from the interstellar
medium \citep{walter1996} has been ruled out by the measurement,
in three of them, of large proper motions that imply an accretion
rate far too small to power the observed luminosity (see, e.g.,
\citet{motch2005} and references therein). Pulsations have been
found in five (or possibly six)  XDINS, with periods in the 3--12
s range, pulsed fractions between  4\% and 18\%, and, in
two cases, period derivatives of the order of 10$^{-13}$ s
s$^{-1}$ \citep{kaplan2005a,kaplan2005b}. These spin-down values
yield characteristic ages of $\sim$1--2 Myrs, spin-down
luminosities of 4--5$\times$10$^{30}$ erg s$^{-1}$, and, with the
canonical dipole radiation braking assumptions, magnetic fields of
a few 10$^{13}$ G. It is thus believed that the X--ray emission in
XDINS is powered by the neutron star cooling.

\rxj\ received particular interest since it is the brightest XDIN,
its X--ray flux does not show any apparent contamination from non
thermal magnetospheric emission, and a parallactic  distance
(117$\pm$12 pc) has been reported \citep{walter2002}. These
properties make \rxj\ an ideal target to derive information on the
neutron star radius and thus constrain the equation of state of
matter at super nuclear density through a detailed modelling of
its surface thermal emission. Quite surprisingly, even the X--ray
spectra with the highest resolution and the best statistics,
obtained in long Chandra and XMM-Newton observations, do not show
any line feature and can be fit by a simple Planckian with
temperature kT=63 eV \citep{burwitz2003}.

Despite extensive searches, no pulsations have ever been detected
in \rxj\ \citep{pons2002,ransom2002,drake2002,burwitz2003}. The
tight upper limits on the pulsed fraction ($<$1.3\% at 2$\sigma$)
obtained for periods between 0.02 and 1000 s \citep{burwitz2003},
led to the speculation that \rxj\ could be a millisecond pulsar,
not yet seen to pulsate owing to the reduced sensitivity for short
periods of the observations carried out up to now.
Recently, an upper limit on the pulsed fraction of 2.1\%
(at 1$\sigma$) was reported for periods in the  range 1-20 ms
\citep{zavlin2006}. Here we report the discovery of pulsations at
7 seconds, obtained thanks to the very high counting statistics
provided by the most recent XMM-Newton observation of \rxj .

\section{Timing analysis and results}

A 70 ks long observation of \rxj\ was carried out with the
XMM-Newton satellite starting on 2006 October 24 at 00:30 UT.  We
used the data of the EPIC instrument (0.1-12 keV), consisting of
two MOS and one pn cameras \citep{turner01,struder01}. All the
cameras mounted the thin filter and were operated in Small Window
mode, yielding  time resolutions of 6 ms and 0.3 s for the pn and
MOS, respectively. We  processed the data using version 7.0 of the
\emph{XMM-Newton Science Analysis System}; the event files were
created using the tasks \emph{epproc} and \emph{emproc} with
default options and filtered to exclude the time intervals of high
particle background, resulting in net (dead-time corrected)
exposure times of 47 ks and 68 ks for the pn and MOS,
respectively.

We started the timing analysis using the data of the pn camera and
selecting the energy range 0.15-1.2 keV. Due to the very soft
spectrum of \rxj , practically no source photons are detected at
higher energies. We used a circular extraction region centered at
the source position and with radius 40$''$. This resulted in
$\sim$356,600 counts, of which $\sim$2,700 can be ascribed to the
background.  The arrival times  were corrected to the Solar System
barycenter.  In order to search for periods of a few seconds, as
seen in the other sources of this class, we rebinned the counts at
$\sim$0.278 s and computed the Fourier power spectrum. A
significant peak in the power spectrum was found at the frequency
of 0.14174 Hz. Taking into account the number of searched periods
(131072), the peak value of 38.46 corresponds to a chance
probability of 6$\times10^{-4}$. To better estimate the period, we
used the epoch folding technique and fitted the peak in the
$\chi^2$ versus trial period distribution as described in
\citet{leahy87}, obtaining P=7.05514$\pm$0.00007 s. The
corresponding folded light curve is shown in the top panel of
Fig.~1. No significant differences in the pulse profiles were
found by dividing the counts in soft and hard energy intervals.
The phase averaged spectrum is well fit with absorbed blackbody
parameters fully consistent with those of the previous XMM-Newton
observations of \rxj\ \citep{haberl2006}. The pulsed fraction,
derived by fitting a sinusoid to the background subtracted pn
light curve, is (1.6$\pm$0.2)\%. This small pulsed fraction
explains why the periodicity was not discovered in previous
observations and is not significantly detected in the MOS data.
The two MOS together collected only $\sim$137,900 counts from a
40$''$ radius extraction region, less than half of the pn ones. It
is thus not surprising that the MOS 0.15-1.2 keV folded  light
curve (bottom panel of Fig.~1) is statistically consistent with a
constant flux.

In order to confirm the periodicity seen in the pn data of
October 2006, we analyzed the other XMM-Newton observations listed
in Table 1. After reducing  the data as described above, we
searched for the 7 s pulsations with the Z$^2$ test
\citep{buccheri83}, which has a good sensitivity for sinusoidal
signals. For each observation, we restricted our search to the
range of periods obtained by a backward extrapolation from the
current value, assuming a conservative period derivative
$|\pdot|<5\times10^{-10}$ s s$^{-1}$. During the April 2004
observation the pn was in Timing mode and the two MOS were in Full
Frame mode with a time resolution of only 2.6 s, that caused
significant photon pile-up in the MOS data. Therefore we used only
the pn data, selecting the energy range 0.28-1.2 keV in order to
avoid the soft flares produced by high-energy particles in the pn
Timing mode \citep{burwitz2004}.  In all the other observations we
used the sum of the 0.15-1.2 keV counts from the pn and the two
MOS\footnote{For the 2002 observation the counts below 0.18 keV in
the MOS1 were excluded in order to eliminate the electronic noise
present in the MOS timing mode}.

The distributions that we obtained for the Z$^2$ statistics
are shown in  Fig.~2, where it can be seen that a maximum
of Z$^2$ is always present at the expected pulse period. To
evaluate the statistical significance of these detections, we must
take into account the number of periods searched in each
observation. The most significant detection is that of April 2002,
where, considering the 335 independent periods in the considered
range (6.98--7.13 s), the peak Z$^2$=32.22 corresponds to a chance
probability of 335$\times e^{-32.22/2}$= 3$\times10^{-5}$. In a
similar way we derived the following chance probabilities for the
other observations: 7$\times10^{-5}$ (April 2004),
5$\times10^{-4}$ (September 2004), 2$\times10^{-3}$  (September
2005), and 2$\times10^{-2}$ (March 2006).

These values indicate that the periodicity is detected in
all the observations, although sometimes with small Z$^2$ values,
as expected for the case of a small pulsed fraction.  The
relatively low significance of the March 2006 detection, despite a
number of counts and duration similar to the October 2006
observation, is not particularly surprising. In fact, with about
half a million counts, a $\sim$1\% modulation has a probability of
~10\% of giving a Z$^2$ as small as the observed value of 15.5
\citep{brazier1994}.

The best periods and pulsed fractions, determined in each
observation as described above for the October 2006 data, are
given in Table~1, while the folded light curves are presented in
Fig.~3. Although this figure seems to suggest some variations in
the pulse profiles, we verified by means of a Kolmogorov-Smirnov
test that, taking into account the unknown relative phase
alignment, all the light curves are fully compatible with the same
profile. We determined the relative phase shifts of the
six observations by fitting sinusoidal functions to the folded
light curves. After applying these shifts, we summed all the pn
and MOS data taken in  Small Window mode to obtain the averaged
light curves shown in Fig.~4 for the total (0.15-1.2 keV), soft
(0.15-0.26 keV) and hard (0.26-1.2 keV) energy ranges. The
corresponding pulsed fractions are (1.17$\pm$0.08)\%,
(0.88$\pm$0.11)\%, and (1.5$\pm$0.11)\%.

All the period measurements, spanning more than four years, are
consistent with an average value of P=7.05515$\pm$0.00004 s. With
a linear fit to the period values we can limit the period
derivative to $-1.2\times10^{-12} < \pdot <1.9\times10^{-12}$ s
s$^{-1}$ (90\% c.l.).

We searched for the \rxj\ pulsations also in the long
Chandra observation performed with the LETG/HRC-S instrument in
October 2001. Our search, based on about 91,600 counts extracted
from the zero-order image, gave a negative result. This is not
surprising, since for this number of counts and a $\sim$1.2\%
pulsed fraction, the expectation value for the Z$^2$ statistics is
only 6.6. Furthermore, the long time span of the observation
yields an intrinsic frequency resolution a factor ~10 better than
that of the best XMM-Newton observation. This means a tenfold
increase in the number of statistically independent periods that
have to be examined and thus a higher threshold for a significant
detection.

\section{Conclusions}

Our discovery of the long sought periodicity in \rxj\    and the
upper limit derived for   $\pdot$  can be used to infer  a
characteristic age $\tau>6\times10^4$ yr, a magnetic field
B$<1.2\times10^{14}$ G, and a spin-down luminosity
$<2\times10^{32}$ erg s$^{-1}$. These values are fully consistent
with those of the other six XDINS.

The pulsed fraction of \rxj\ is significantly smaller than that of
the other members of the group, which have pulsed fractions in the
range from 4\% to 18\%. The XMM-Newton data give also
evidence for a pulsed fraction increasing as a function of the
energy. The small pulsed fraction in \rxj\ might simply result
from geometrical orientation effects or by a particularly small
gradient in the surface temperature distribution. Detailed
phase-resolved spectral modelling, that might discriminate between
the different possibilities, will be reported elsewhere. Here we
simply  note that the October 2006 pn spectra for  the maximum and
minimum of the folded light curve are both consistent with the
best-fit absorbed blackbody model of the phase-averaged spectrum,
simply re-scaled in normalization. Letting all the fit parameters
free to vary, the maximum allowed difference in the temperatures
of the two spectra  is of  kT$_{\rm BB}$=1.2 eV (at the 90\%
c.l.). Repeating the same exercise for 5 phase bins, we find again
no significant spectral variations, with all the parameters within
the following ranges: N$_{\rm H}=$(4--8)$\times$10$^{19}$
cm$^{-2}$ and kT$_{\rm BB}$=61.4--63.2 eV.

Our results indicate that, except for the striking absence of
deviations from a pure blackbody in its X--ray spectrum,  \rxj\
shares most of the properties of the other members of the XDINS
group. More importantly, the  discovery  of  pulsations opens the
way for the determination of the timing-dependent parameters that
are crucial for a more detailed modelling of the star's surface
emission properties.

 \acknowledgments

Based on observations obtained with XMM-Newton, an ESA science
mission with instruments and contributions directly funded by ESA
Member States and NASA. This work has been  supported by the
Italian Space Agency under contract ASI/INAF/I/023/05/0.

\acknowledgments

\clearpage

\begin{table}
\caption{Log of \xmm\ observations of \rxj. The exposure
times take into account the filtering for high background time
intervals and the instrumental dead-time. The time resolution of
the small window (SW) mode is 6 ms for the pn and  0.3 s for the
MOS, while for the Timing (TI) mode it is 0.03 ms for pn and 1.5
ms for MOS. The period values are obtained with the sum of pn and
MOS counts. The pulsed fractions are computed from sinusoidal fits
of the background subtracted pulse profiles (PF=A/C, where the
count rate is given by C+Asin($\phi-\phi_0$)). Errors are at
1$\sigma$.} \smallskip
\begin{tabular}[c]{ccccrcc}
\hline \hline
 Date & Instr. & Mode   & Exp.  & Counts & Period & PF \\
   & & & (ks) & & (s) & (\%)\\
\hline \hline
2002-04-08&pn&SW&40&304,052 &7.05510$\pm$0.00013&1.4$\pm$0.2\\
&MOS1&TI&57&53,529&&1.1$\pm$0.5\\
&MOS2&SW&56&82,837& &0.8$\pm$0.4\\
\hline
2004-04-17&pn&TI&31&109,232& 7.05502$\pm$0.00012&2.4$\pm$0.3\\
\hline
2004-09-24&pn&SW&23&178,718&7.05524$\pm$0.00010&1.1$\pm$0.2\\
&MOS1&SW&68&46,284&&2.0$\pm$0.5\\
&MOS2&SW&68&52,865& &2.0$\pm$0.5\\
\hline
2005-09-24&pn&SW&23&169,803&7.05518$\pm$0.00027&1.6$\pm$0.2\\
&MOS1&SW&27&26,936&&0.8$\pm$0.6\\
&MOS2&SW&27&29,706& &1.5$\pm$0.6\\
\hline
2006-03-26&pn&SW&49&365,050&7.05526$\pm$0.00016&0.8$\pm$0.2\\
&MOS1&SW&67&64,895&&1.4$\pm$0.4\\
&MOS2&SW&67&74,821& &0.7$\pm$0.4\\
\hline
2006-10-24&pn&SW&47&356,605&7.05514$\pm$0.00007&1.6$\pm$0.2\\
&MOS1&SW&68&65,941&&1.1$\pm$0.4\\
&MOS2&SW&68&71,973& &1.1$\pm$0.4\\
\hline

\end{tabular}
\label{tab:axp_obs}
\end{table}

\clearpage

\begin{figure}
\epsscale{.7}
\plotone{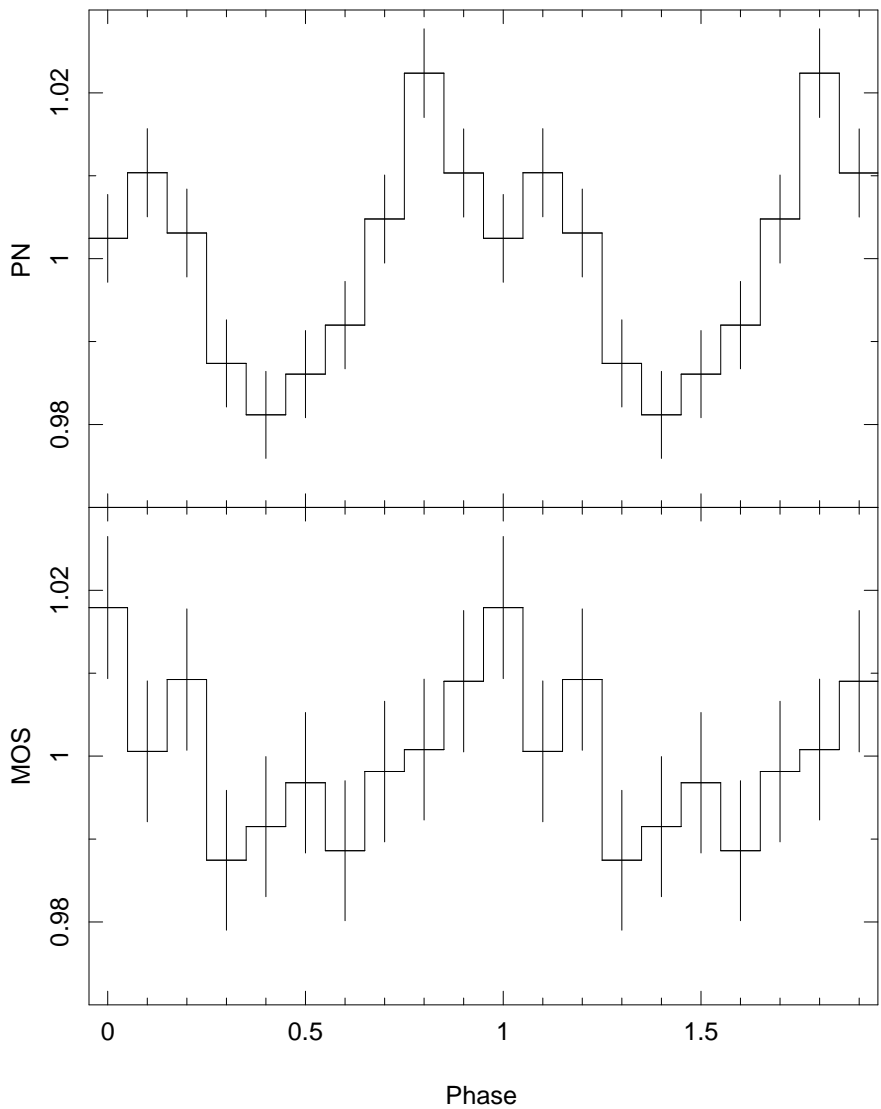}
\caption{Folded light curves of
\rxj\ in the 0.15-1.2 keV energy
 range obtained in the October 2006
 XMM-Newton observation. Top: EPIC pn, Bottom: EPIC MOS1+MOS2.
Fitting the  light curves with a constant value gives
$\chi^2$ values of 56.8 (pn) and 8.6 (MOS1+MOS2) for 9 d.o.f.,
corresponding to chance probabilities of 6$\times10^{-9}$ and
0.47, respectively.
 \label{f1}}
\end{figure}

\clearpage

\begin{figure}
\epsscale{1.0}
\plotone{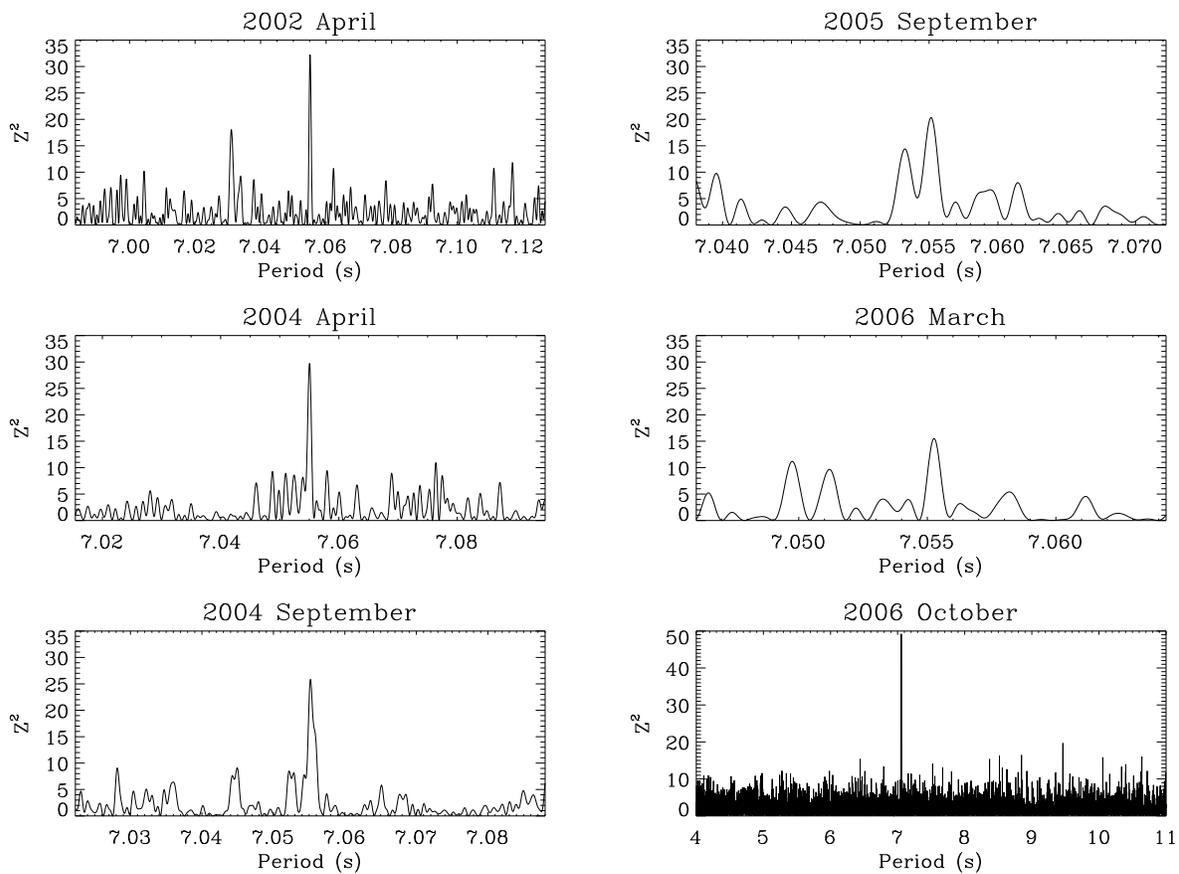}
\caption{Distributions of the
Z$^2$ statistics for the XMM-Newton observations
 of \rxj . The period ranges shown in the first five panels correspond to those used in
 the period search and derived from a conservative $\pdot$ assumption (see text for details).
\label{f2}}
\end{figure}

\clearpage

\begin{figure}
\plotone{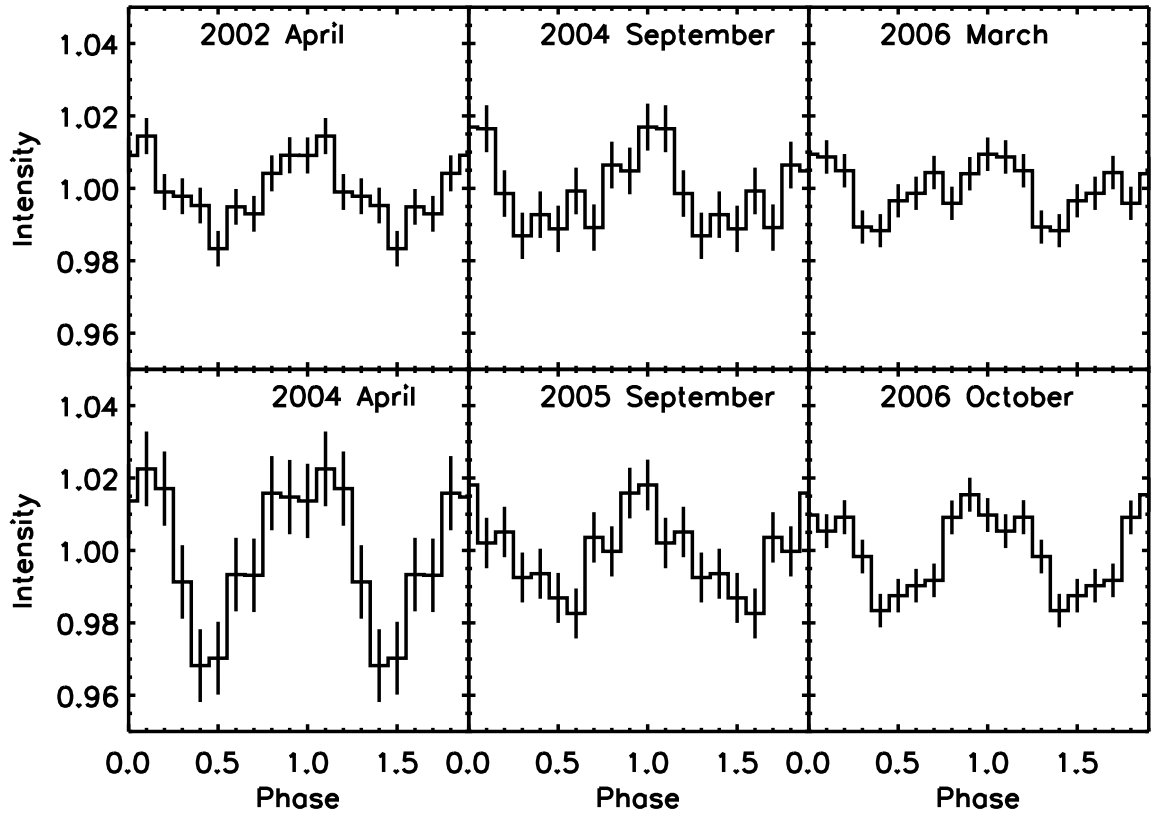}
\caption{Background subtracted light curves of
\rxj\ folded at the best-fit   periods (see Table~1). The light
curves are normalized to the average count rate. The phase
alignment between the different panels is arbitrary. All the light
curves are from the sum of pn and MOS counts in the 0.15-1.2 keV
range,  except for the 2004 April observation where only the
0.28-1.2 keV pn counts have been used. \label{f3}}
\end{figure}

\clearpage

\begin{figure}
\plotone{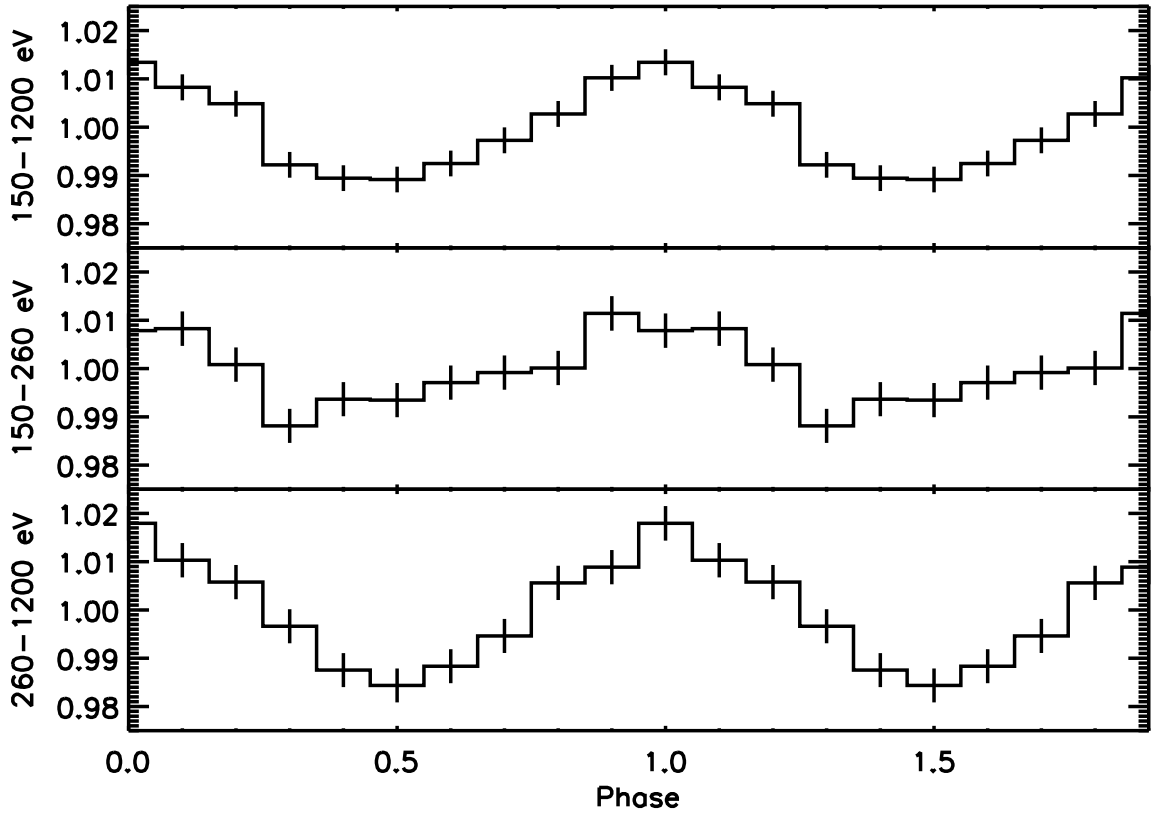}
\caption{Light curves of \rxj\ in the total
(0.15-1.2 keV), soft (0.15-0.26 keV), and hard (0.26-1.2 keV)
energy ranges obtained by summing all the pn and MOS data in small
window mode. \label{f4}}
\end{figure}

\end{document}